\documentclass[preprint,aps,prc,superscriptaddress,showkeys]{revtex4-2}
\usepackage{graphicx} 
\usepackage{amsmath}
\usepackage{appendix}

\begin{document}

\title{Nuclear shapes of Nb isotopes}


\author{Esperanza Maya-Barbecho}
\email{esperanza.maya@dci.uhu.es}
\affiliation{Departamento de Ciencias Integradas y Centro de Estudios Avanzados en F\'isica, Matem\'atica y Computaci\'on, Universidad de Huelva, 21071 Huelva, Spain}

\author{Jos\'e-Enrique Garc\'ia-Ramos}
\email{enrique.ramos@dfaie.uhu.es}
\affiliation{Departamento de Ciencias Integradas y Centro de Estudios Avanzados en F\'isica, Matem\'atica y Computaci\'on, Universidad de Huelva, 21071 Huelva, Spain}
\affiliation{Instituto Carlos I de F\'{\i}sica Te\'orica y Computacional, Universidad de Granada, Fuentenueva s/n, 18071 Granada, Spain}

\begin{abstract}
\begin{description}
\item[Background]  
The study of the structure of odd-mass nuclei in regions characterized by the interplay of multiple particle-hole configurations represents a major challenge in nuclear structure physics. The odd-mass niobium isotopes ($Z = 41$), located near the $N = 60$ region, are of particular interest due to the occurrence of shape coexistence and quantum phase transitions.

\item[Purpose]  
This work aims to investigate the structure of the $^{93-103}$Nb isotopes using the intrinsic-frame formalism of the interacting boson–fermion model with configuration mixing (IBFM-CM). The goal is to determine the nuclear shapes and explore the phenomena of shape coexistence, configuration crossing, and quantum phase transitions.

\item[Method]  
We employ the intrinsic formalism of the IBFM-CM, which includes both 0p–0h (regular) and 2p–2h (intruder) configurations interacting with the unpaired nucleon. This approach provides a self-consistent framework to study energy surfaces, shape coexistence, and intruder bands for both positive- and negative-parity states. A realistic Hamiltonian for niobium, determined in previous studies, is adopted.

\item[Results]  
The formalism is applied to the $^{93-103}$Nb isotopes for both positive- and negative-parity bands. A detailed analysis of the mean-field energy surfaces has been performed, including axial energy curves, triaxial energy surfaces in the $\beta–\gamma$ plane, and the corresponding equilibrium deformation parameters. The results reveal clear evidence of configuration coexistence and crossing along the isotopic chain.

\item[Conclusions]  
We have applied the recently developed intrinsic-state formalism of the IBFM-CM using a realistic Hamiltonian for a chain of niobium isotopes. The existence of crossing configurations has been demonstrated around $N = 60$, corresponding to a quantum phase transition previously identified in the Sr and Zr isotopic chains. Furthermore, we find that the presence of an unpaired nucleon in Nb influences the abruptness of the quantum phase transition, underscoring the sensitivity of the structural evolution to single-particle degrees of freedom.
\end{description}
\end{abstract}

\date{\today}
\maketitle

\section{Introduction}
\label{sec-intro}
The emergence of shape coexistence and the rapid onset of nuclear deformation are among the most remarkable structural phenomena observed in atomic nuclei \cite{hey83, wood92,heyde11,Garr22}. These effects are particularly significant in the mass region $A \sim 100 $, where a sudden transition from near-spherical to well-deformed shapes is found around neutron number $N=60$ \cite{Togashi16,Garc19,Gavr19}. This transition, caused by the interplay of proton and neutron orbits near shell or sub-shell closures, is well documented in nuclei such as Sr, Zr, and Mo, and has been associated with shape coexistence and quantum phase transitions (QPTs) \cite{Garc20,Maya2023}. The standard explanation for the rapid onset of deformation in this mass region is based on the interplay between the $1g_{9/2}$ orbital for protons and the $1g_{7/2}$ for neutrons, which enhance the deformation around $N=60$ and beyond. This approach, as suggested by Federman and Pittel in 1979 \cite{Fede79a,Fede79b}, emphasizes the importance of the simultaneous occupation of neutrons and protons of spin-orbit partners to favor the appearance of deformation. A similar explanation was proposed by Otsuka and collaborators in \cite{Togashi16} in terms of the tensor component of the nuclear force. 

Odd-mass nuclei in this region, and in particular the niobium isotopes ($Z=41$), provide an ideal testing ground to study how the presence of an unpaired nucleon modifies the nuclear structure. The Nb isotopes are characterized by a single unpaired proton, occupying the single-particle orbits $1g_{9/2}$, $1f_{5/2}, 2p_{3/2},$ and $2p_{1/2}$, which couples to an even-even core. This coupling introduces additional complexity into the structural evolution, offering an opportunity to explore configuration mixing, shape coexistence, and the enhancement or suppression of QPTs in odd-even nuclei. As a matter of fact, the presence of the odd nucleon in $A\approx 100$ mass area enhance the onset of deformation as shown in Ref.\cite{Maya2025}.

In order to study such odd-even nuclei, we will use the interacting boson-fermion model (IBFM) \cite{Iach1991} as a natural extension of the interacting boson model (IBM) \cite{iach87} that incorporates the degrees of freedom of a single unpaired nucleon. When the model space is enlarged to include both regular $(0p-0h)$ and intruder ($2p–2h$) configurations across a shell gap \cite{duval81,duval82}, one obtain the IBM with configuration mixing (IBM-CM). In an analogous way, in the case of the IBFM also two configurations can be included resulting in the IBFM with configuration mixing (IBFM-CM). This framework becomes a powerful tool to describe shape coexistence and configuration crossings in odd-mass nuclei.

While previously Nb isotopes have been studied in the laboratory frame to describe spectroscopic properties \cite{Gavr22b,Gavr2023}, a more visual and geometric understanding can be gained through a mean-field approach. Moreover, recently it has been proposed a new intrinsic-state formalism for the IBFM-CM \cite{Maya2025,Levi2025,Maya2025b}. This framework enables the calculation of energy surfaces for odd-mass nuclei, including both axial and triaxial degrees of freedom, and allows the identification of configuration mixing effects and QPTs.

In this work, we apply the IBFM-CM intrinsic-state formalism to the chain of odd-mass Nb isotopes, from $^{93}\mathrm{Nb}$ to $^{103}\mathrm{Nb}$, building upon previous analyses carried out in the laboratory frame \cite{Gavr22b,Gavr2023}, where a realistic Hamiltonian constrained by spectroscopic data was obtained. We explore the evolution of the equilibrium deformation parameters $\beta$ and $\gamma$, analyze axial and triaxial energy surfaces, and investigate the microscopic origin of the sudden onset of deformation at $N=60$. Our results reveal that the presence of the odd proton not only affects the collective behavior of the core but can also sharpen the structural transition already present in the even system, providing enhanced sensitivity to configuration mixing effects.

The paper is organized as follows. In Sec.~\ref{sec:formalism}, we summarize the IBFM-CM Hamiltonian and outline the intrinsic-state formalism. The main results, including deformation systematics, axial and triaxial energy surfaces, and configuration mixing effects, are presented and discussed in Sec.~\ref{sec:results}. In Sec.~\ref{sec:qpt}, a discussion of QPTs is presented. Finally, conclusions are drawn in Sec.~\ref{sec:conclusions}.

\section{Formalism}
\label{sec:formalism}
\subsection{The Hamiltonian}
The IBFM \cite{Iach1991} extends the IBM \cite{iach87} by coupling a single unpaired nucleon (fermion) to an even-even bosonic core composed of $s$ ($L=0$) and $d$ ($L=2$) bosons. The number of bosons $N$ corresponds to half the number of valence nucleons, excluding the odd fermion.

The general IBFM Hamiltonian can be decomposed as
\begin{equation}
\hat H = \hat H_B + \hat H_F + \hat H_{BF},
\end{equation}
where $\hat H_B$ describes the bosonic core, $\hat H_F$ the single-particle fermion, and $\hat H_{BF}$ the boson-fermion interaction.

To incorporate intruder excitations (2p–2h), the space is extended to a configuration-mixed basis $[N] \oplus [N+2]$, and, hence, the Hamiltonian is enlarged in the following way
\begin{equation}
\hat H = \hat P_N^\dagger \hat H^N \hat P_N + \hat P_{N+2}^\dagger (\hat H^{N+2} + \Delta^{N+2}) \hat P_{N+2} + \hat V_{\text{mix}}^{N,N+2},
\end{equation}
where $\hat P_N$ and $\hat P_{N+2}$ are the projectors onto regular and intruder spaces, respectively. The term $\Delta^{N+2}$ accounts for the energy offset of the intruder configuration, and $\hat V_{\text{mix}}^{N,N+2}$ is the part of the Hamiltonian that connects the two configurations $[N]$ and $[N+2]$.

The mixing term is given by
\begin{equation}
\label{Vmix_boson}
\hat V_{\text{mix}}^{N,N+2} = \omega_0^{N,N+2}[s^\dagger s^\dagger + ss]^{(0)} + \omega_2^{N,N+2}[d^\dagger d^\dagger + \tilde{d} \tilde{d}]^{(0)},
\end{equation}
where it is assumed that the fermion contribution does not depend on the orbit and, hence, it can be absorbed into the boson one. Moreover, it is commonly assumed $\omega_0^{N,N+2} = \omega_2^{N,N+2} \equiv \omega$ (see \cite{Gavr22b,Gavr2023}).

The fermion Hamiltonian is diagonal,
\begin{equation}
\hat H_F^i = \sum_j \epsilon_j^i \hat n_j,
\end{equation}
where $\hat n_j$ is the fermion number operator in orbit $j$, and $\epsilon_j^i$ are the single-particle energies, with $i$ standing for $N$ or $N+2$.

The boson-fermion interaction is structured into three components
\begin{equation}
\hat H_{BF}^i = \hat H_{\text{MON}}^i + \hat H_{\text{QUAD}}^i + \hat H_{\text{EXC}}^i,
\end{equation}
corresponding to the monopole, quadrupole, and exchange terms, respectively (see \cite{Gavr22b,Gavr2023}).

\subsection{Intrinsic-state Formalism}
To explore the geometry of the IBFM-CM Hamiltonian, we employ the intrinsic-state formalism. The first step is to construct a boson condensate that incorporates the deformation parameters $(\beta,\gamma)$\cite{gino80,diep80a,diep80b},
\begin{equation}
|N;\beta,\gamma\rangle = \frac{1}{\sqrt{N!}} \left( \frac{ s^\dagger + \beta \cos\gamma\, d_0^\dagger + \frac{1}{\sqrt{2}} \beta \sin\gamma (d_{+2}^\dagger + d_{-2}^\dagger) }{ \sqrt{1 + \beta^2} } \right)^N |0\rangle.
\end{equation}

To take into account the fermion, we consider that we work, implicitly, with the basis $|N;\beta,\gamma\rangle \otimes |j m\rangle$\cite{Levi1988}.
The IBFM Hamiltonian can then be mapped into the matrix form
\begin{equation}
\mathcal{H}_{\text{IBFM}}(N,\beta,\gamma) = \sum_{j_1 m_1 j_2 m_2} \mathcal{M}_{j_1 m_1, j_2 m_2}(N,\beta,\gamma)\, a_{j_1 m_1}^\dagger a_{j_2 m_2},
\end{equation}
where the matrix elements $\mathcal{M}$ contain contributions from the boson energy surface $E_B(N,\beta,\gamma)$, single-particle energies $\epsilon_j$, and the boson-fermion interaction $E_{BF}(N,\beta,\gamma)$ (see Refs.~\cite{Levi1988, Alon1992, Maya2025b}).
In the axial symmetry case, 
this matrix is block-diagonal in $m$, simplifying the diagonalization. 

The next step is to enlarge the Hilbert space to consider the 2p-2h configuration \cite{Frank02,Frank04,Frank06}, 
\begin{equation}
\label{HH_CM}
\mathcal{H}_{\text{IBFM}}^{\text{CM}}(N,\beta,\gamma) = \sum_{i i';j_1 m_1 j_2 m_2} \mathcal{M}^{CM}_{i i', j_1 m_1, j_2 m_2}(N,\beta,\gamma)\, \hat P_i^\dagger \, a_{j_1 m_1}^\dagger a_{j_2 m_2}\, \hat P_{i'},
\end{equation}
where $i,i'$ can take the values $N$ and $N+2$, corresponding to the regular and intruder subspaces, respectively and,
 
\begin{equation}
\label{H_CM}
{\cal M}^{CM}(N,\beta,\gamma)=\left (
\begin{array}{@{\hspace{0.2em}}c@{\hspace{0.3em}}c@{\hspace{0.2em}}}
{\cal M}(N,\beta,\gamma)& \Omega_{BF}(\beta)\\
\Omega_{BF}(\beta)& {\cal M}(N+2,\beta,\gamma)
\end{array}
\right ),
\end{equation}
with ${\cal M}(N,\beta,\gamma)$ and ${\cal M}(N+2,\beta,\gamma)$ corresponding to the matrices of the regular and the intruder Hamiltonian and $\Omega_{BF}(\beta)$ being the matrix form of the mixing operator $V_\text{mix}(N, \beta)$ written as
\begin{equation}
\label{V_mix}
   V_\text{mix}(N,\beta)=\sum_{i\neq i',j_1, m_1, j_2, m_2} \Omega_{BF}(N,\beta)_{j_1, m_1, j_2, m_2}\; \hat P_i^\dagger \,a^\dagger_{j_1,m_1} a_{j_2,m_2} \hat P_{i'}.
\end{equation}
As a matter of fact, $\Omega_{BF}(N,\beta)$ is diagonal and if Eq.~(\ref{Vmix_boson}) is assumed, it has the same single value over the whole diagonal \cite{Maya2025b}. 

The total dimension of (\ref{H_CM}) is at maximum $\sum_j (2j+1)$ \cite{Maya2025}. The wave functions obtained from the diagonalization of Hamiltonian (\ref{HH_CM}) have the form,
\begin{equation}
\label{eq:wf}
\begin{aligned}    
    |\phi (k; \beta, \gamma)\rangle &= \sum_{j,m} a^{k}_{N; j,m}(\beta,\gamma) |N;\beta,\gamma\rangle\otimes|j m\rangle\\ 
    &+\sum_{j,m} b^{k}_{N+2; j,m}(\beta,\gamma) |N;\beta,\gamma\rangle\otimes|j m\rangle.
\end{aligned}
\end{equation}
Note that the wave functions depends on $\beta$ and $\gamma$. It is also convenient to define the weight of the wave function contained within the [N]-boson subspace, i.e., the sum of the squared amplitudes (similarly it can be defined for the [N+2] sector),
\begin{equation}
\label{eq:reg_content}
    w^k(N;\beta,\gamma) =\sum_{j,m} |a^{k}_{N; j,m}(\beta,\gamma)|^2.
\end{equation}

For each of the eigenvalues, it is needed to obtain the equilibrium values of $\beta$ and $\gamma$, $\beta_0$ and $\gamma_0$, which minimize the corresponding energy surface \cite{Petr2011}. Those equilibrium values are, in principle, different among them and different from the boson ones. This fact is very relevant because the wave functions (\ref{eq:wf}) corresponding to the equilibrium values will lose their orthogonal character, although it is recovered in the case of axial symmetry where the projection is preserved. 

\section{Results}
\label{sec:results}
The isotope chain  $^{93-103}$Nb was previously described in \cite{Gavr22b,Gavr2023} 
were the value of the Hamiltonian parameters were obtained by performing a complete analysis in the framework of the laboratory frame of the IBFM-CM. In this work, we will consider those parameters, in order to elaborate an study of the same isotope chain using the IBFM-CM intrinsic-state formalism.

Within this framework, we will consider a boson core coupled to an unpaired proton located in the $1f_{5/2}, 2p_{3/2}, 2p_{1/2}$ orbits for negative parities, while in the $1g_{9/2}$ shell for positive  parities. This offers the opportunity to test the proposed formalism in a complete realistic case in both single and multiple-j cases.

The bosonic part of the Hamiltonian in each configuration, $i=[N]$ and $[N+2]$, corresponds to the extended consistent-Q formalism \cite{warner83}
\begin{equation}
\hat H_B^i = \varepsilon_i \hat n_d + \kappa_i' \hat L \cdot \hat L + \kappa_i \hat Q(\chi_i) \cdot \hat Q(\chi_i),
\end{equation}
where $\hat n_d$ is the $d$-boson number operator, $\hat L$ the angular momentum operator, and $\hat Q(\chi_i)$ the quadrupole operator parameterized by $\chi_i$. This form provides sufficient flexibility to describe a wide range of nuclei.

Concerning the boson-fermion interaction, we will follow the most common approach \cite{Iach1991} assuming that it is composed of three components, namely, the monopole, the quadrupole and the exchange terms,
\begin{equation}  \hat{H}_{\mathrm{BF}}^i=\hat{H}_{\mathrm{BF}}^{\mathrm{MON},i}+\hat{H}_{\mathrm{BF}}^{\mathrm{QUAD},i}+\hat{H}_{\mathrm{BF}}^{\mathrm{EXC},i}
\end{equation}
where,
\begin{equation}
  \begin{aligned}
    & H_{\mathrm{BF}}^{\mathrm{MON},i}=\sum_j A_j^i\left[\left[d^{\dagger} \times \tilde{d}\right]^{(0)} \times\left[a_j^{\dagger} \times \tilde{a}_j\right]^{(0)}\right]_0^{(0)}, \\
    & H_{\mathrm{BF}}^{\mathrm{QUAD},i}=\sum_{j j^{\prime}} \Gamma_{j j^{\prime}}^i\left[\hat{Q}(\chi_i) \cdot\left[a_j^{\dagger} \times \tilde{a}_{j^{\prime}}\right]^{(2)}\right]_0^{(0)}, \\
    & H_{\mathrm{BF}}^{\mathrm{EXC},i}= \sum_{j j^{\prime} j^{\prime \prime}} \Lambda_{j j^{\prime}}^{j^{\prime \prime},i}:\left[\left[d^{\dagger} \times \tilde{a}_j\right]^{\left(j^{\prime \prime}\right)} \times\left[\tilde{d} \times a_{j^{\prime}}^{\dagger}\right]^{\left(j^{\prime \prime}\right)}\right]_0^{(0)}:,
\end{aligned}
\end{equation}
$:$ stands for normal order and $\tilde{a}_{j,m}= (-1)^{j-m} a_{j,-m}$.

Moreover, we closely follow the microscopic interpretation of the IBFM proposed in \cite{Schol1985}. Hence,
\begin{equation}
\begin{aligned}
A_j & =-\sqrt{5(2 j+1)} A_0, \\
\Gamma_{j j^{\prime}} & =\sqrt{5} \gamma_{j j^{\prime}} \Gamma_0, \\
\Lambda_{j j^{\prime}}^{ j^{\prime \prime}} & =-2 \sqrt{\frac{5}{2 j^{\prime \prime}+1}} \beta_{j j^{\prime \prime}} \beta_{j^{\prime} j^{\prime \prime}} \Lambda_0,
\end{aligned}
\end{equation}
where
\begin{equation}
    \begin{aligned}
\gamma_{j j^{\prime}} & =\left(u_j u_{j^{\prime}}-v_j v_{j^{\prime}}\right) Q_{j j^{\prime}}, \\
\beta_{j j^{\prime}} & =\left(u_j v_{j^{\prime}}+v_j u_{j^{\prime}}\right) Q_{j j^{\prime}}, \\
Q_{j j^{\prime}} & =\left\langle j|| Y^{(2)} \| j^{\prime}\right\rangle,
\end{aligned}
\end{equation}
being $Y^{(2)}$ the spherical harmonic. 

The coefficients $u_j$ and $v_j$ are obtained conducting a BCS calculation over the single particle orbits $1g_{9/2}$, $1f_{5/2}, 2p_{3/2}$, and $2p_{1/2}$, obtaining additionally the single quasi-particle energies. 

The boson-fermion mixing term is assumed to have a structure similar to that of the boson one,
\begin{equation}
  \hat{V}_{\rm BF, mix}^{N,N+2}=\sum_j\hat{n}_j\omega_j\big([s^\dag\times s^\dag + s\times s]^{(0)} + [d^\dag\times d^\dag+\tilde{d}\times \tilde{d}]^{(0)}\big).
\label{eq:vmix_fermion}
\end{equation}
In this case, $\omega_j$ is considered independent on $j$ and thus its effect is absorbed into the boson mixing term, reducing the number of free parameters and obtaining Eq.~(\ref{Vmix_boson}). 

In \cite{Gavr22b,Gavr2023}, it can be found all the coefficients appearing in the Hamiltonian and in the transition operators. Those coefficients have been obtained trough a least-squares fit to the available experimental information.  
\begin{figure}[hbt]
    \centering
    \includegraphics[width=0.6\textwidth]{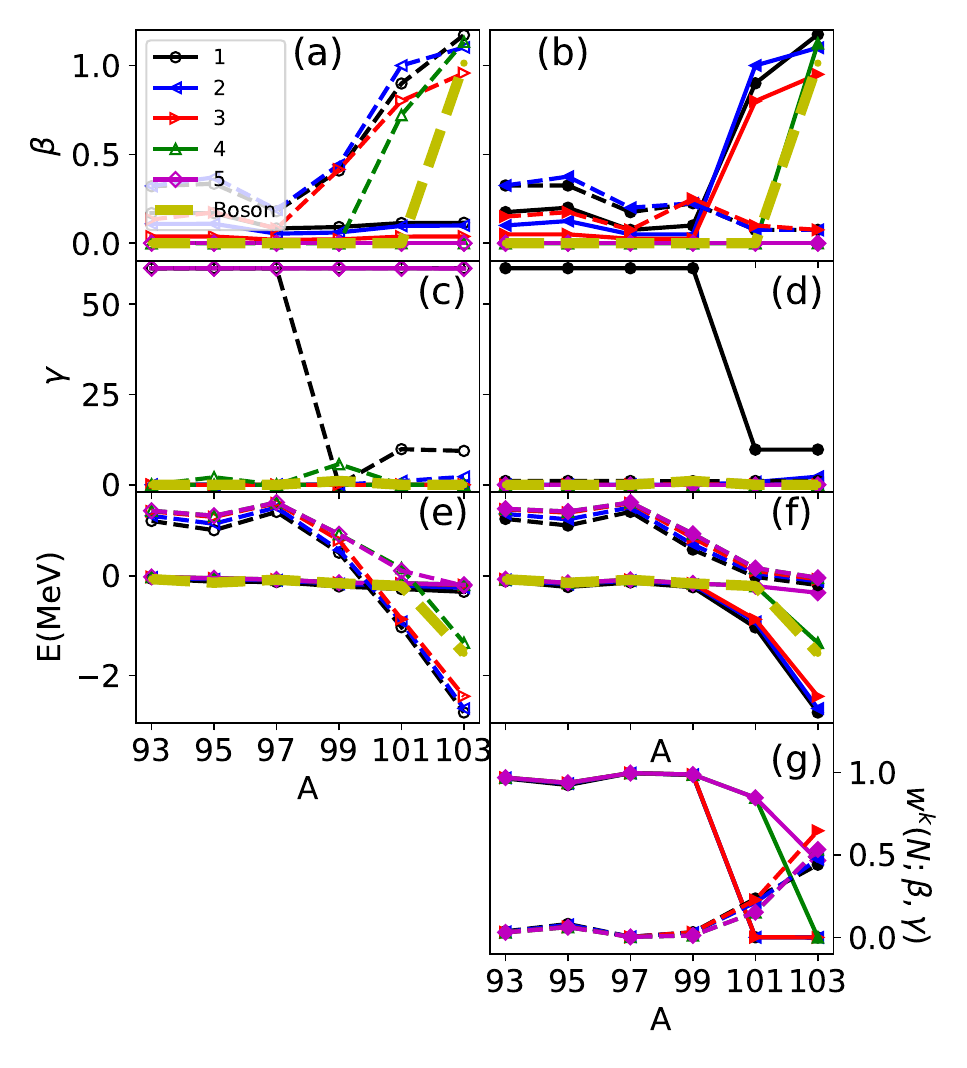}
    \caption{Equilibrium deformation values, excitation energies, and regular content of the wavefunction for the positive parity states. (a) equilibrium values of $\beta$ for the unmixed case ($V_{mix}=0$), full lines for the regular states and dashed for the intruder ones, thick dashed yellow line for the IBM-CM result; (b) equilibrium value of $\beta$ for $V_{mix}\neq 0$, full lines for the first five states and dashed for the last five ones; (c) equilibrium value of $\gamma$ for $V_{mix}=0$; (d) equilibrium value of $\gamma$ for $V_{mix}\neq 0$;  (e) excitation energies for $V_{mix}=0$; (f) excitation energies for $V_{mix}\neq 0$; (g) regular content of the wave function for $V_{mix}\neq 0$.}
    \label{fig-energies-betas-gammas-pos}
\end{figure}

\subsection{Positive parity states}
\label{sec:positive}
We begin by analyzing the case in which the odd proton occupies the $1g_{9/2}$ orbit. Under these conditions, the Hamiltonian matrix to be diagonalized has dimension 10, which arises from simultaneously including both regular and intruder configurations.  

The first step in the analysis is to determine the equilibrium values of the deformation parameters. To clarify the role played by the mixing between regular and intruder configurations, it is useful to examine two different situations: first, the case without configuration mixing ($V_{\rm mix}=0$), and then the case in which the mixing interaction is included ($V_{\rm mix}\neq 0$).
\begin{figure}
    \centering
    \includegraphics[width=0.5\textwidth]{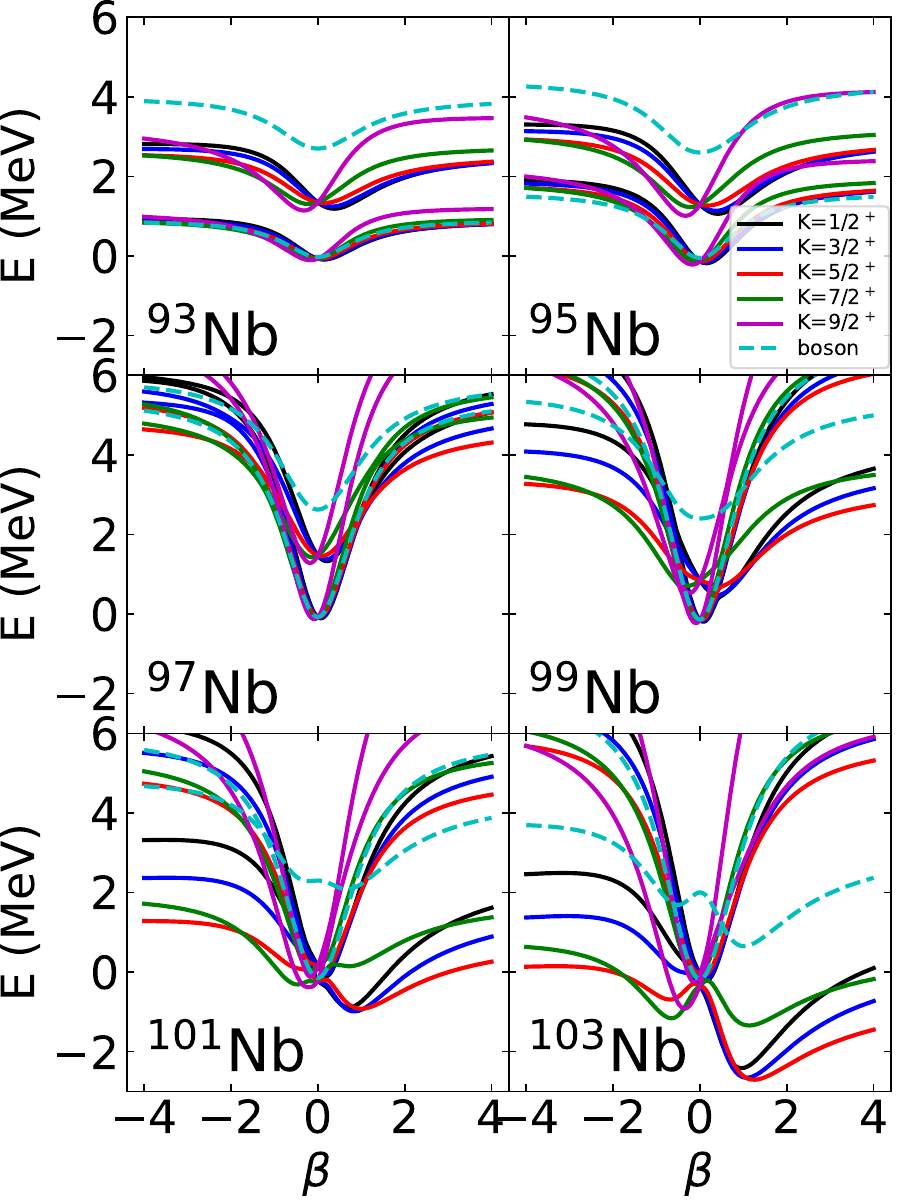}
    \caption{Axial energy curves as a function of $K^\pi$ for positive-parity states with $V_{mix}\neq 0$. For reference, the unmixed regular and intruder energy surfaces are also included.}
    \label{fig-2D-energies-pos}
\end{figure}

The unmixed results are shown in the left column of Fig.~\ref{fig-energies-betas-gammas-pos}. In this situation, the regular (solid lines) and intruder (dashed lines) configurations are clearly separated, both in their excitation energies [panel (e)] and in their equilibrium deformations [panels (a) and (c)]. The figure displays the eigenvalues obtained for each configuration in panel (e): the first five lines correspond to the regular sector, while the next five belong to the intruder sector. The thick dashed yellow line indicates the energy reference associated with the boson core, i.e., the IBM-CM result. For each state, the values of the deformation parameters $\beta$ and $\gamma$ that minimize the energy are also included in panel (a) and (c) respectively.   

An important aspect of these results is the crossing between the regular and intruder configurations, which exhibit very different equilibrium deformations. As the neutron number increases, the intruder configuration gradually becomes lower in energy, relative to the regular one. This behavior provides an explanation for the sudden onset of deformation observed along the isotopic chain. 

The values obtained for the equilibrium deformation parameters also point towards this direction. Intruder states are associated in panel (a) with larger $\beta$ values, which indicates a stronger deformation. This becomes especially relevant from $A=99$ onwards, where the intruder configuration starts to dominate the low-energy spectrum. In addition, for the heaviest isotopes ($A=101$ and $103$), the calculated $\gamma$ values are significantly different from $0^\circ$ or $60^\circ$, pointing into the development of triaxial shapes in this region. 

Since configuration mixing is not included in this calculation, the regular and intruder sets of states remain well separated, which allows us to clearly identify the point where the crossing of configurations occurs and the intruder configuration becomes the ground state of the system, between $A=99$ and $A=101$.

When configuration mixing is included ($V_{\rm mix}\neq 0$), shown in the right column of Fig.~\ref{fig-energies-betas-gammas-pos}, the spectrum exhibits a more intricate structure regarding regular and intruder states, since they cannot be distinguished as in the previous case. The five lowest eigenvalues are now plotted with solid lines, while the next five are shown with dashed lines. 

The general behavior of the excitation energies and of the deformation parameters $\beta$ and $\gamma$ remains qualitatively similar to the unmixed situation. However, because regular and intruder configurations are now mixed, the wave functions become fragmented among the two configurations. In panel (g), it is depicted the fraction of the wave function in the regular sector (\ref{eq:reg_content}), $w^k(N;\beta_0,\gamma_0)$. 
Here, it is easily observed how for $^{93-99}$Nb the lowest states have a  pure regular character while intruder the upper ones. This situation changes when the states start crossing, in particular for the first, second and third states (black, blue and red lines) in $^{101}$Nb and, finally, for the fourth  one (green line) in $^{103}$Nb. It is remarkable that the observed behavior for the ground state coincides with the laboratory frame calculation presented in Fig.~19 of Ref.~\cite{Gavr2023}.
As extra neutrons are added, this distinction gradually fades, and the wave functions become fragmented for the high lying states and them become strongly mixed. 
An additional consequence of the crossing is that the regular content of the states is reduced and, as a matter of fact, the sum of $\omega^k(N;\beta_0,\gamma_0)$ over all the states is smaller than half the number of states, as it should be if orthogonality were preserved. In other words, the regular content of the states is lower than it should be because of the non-orthogonality of the states due to large difference of deformation between the lowest and highest states. This fact suggests that in the case of the heaviest isotopes there is a larger presence of intruder states in the low-lying spectrum.
\begin{figure}[hbt]
    \centering
    \includegraphics[width=0.5\textwidth]{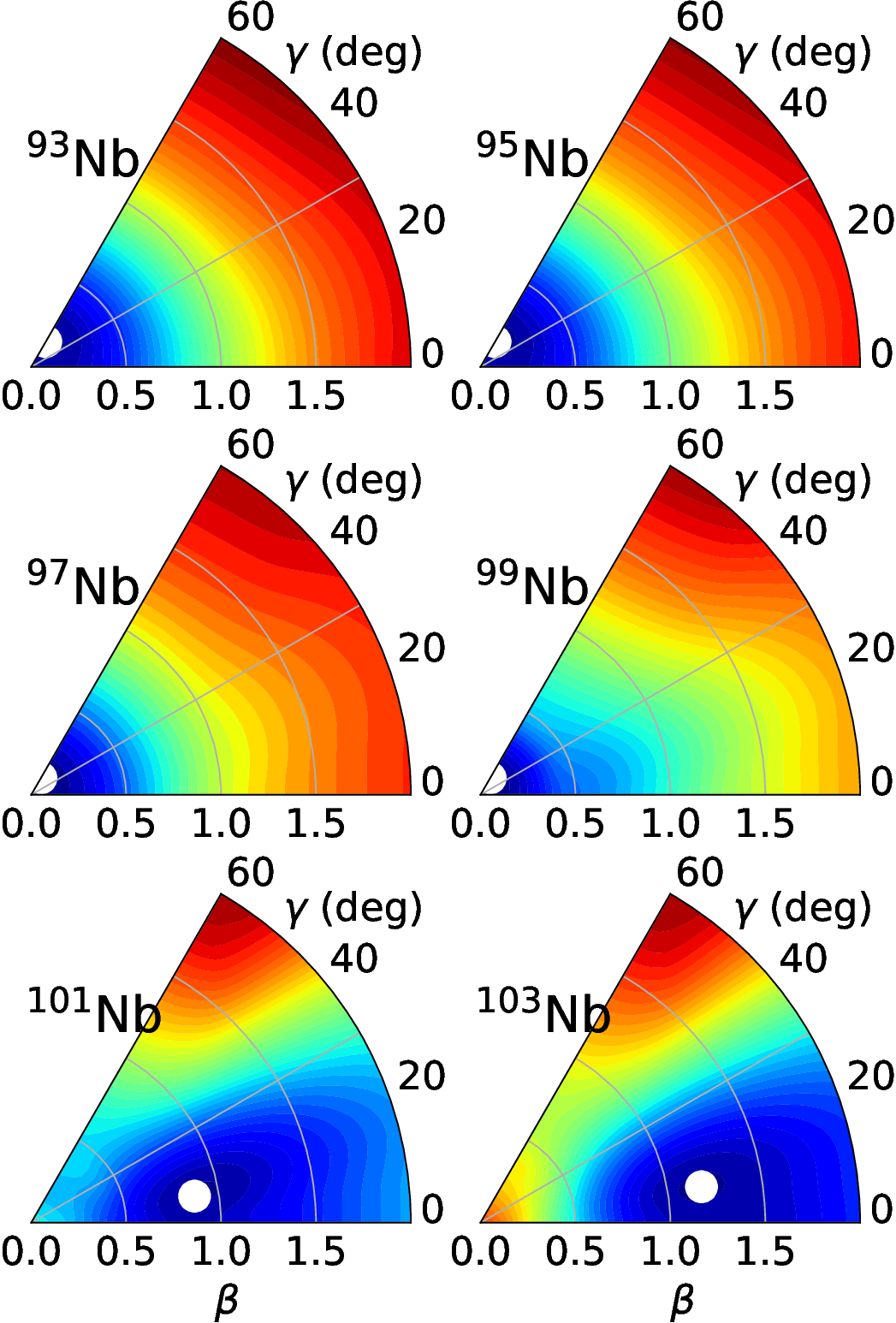}
    \caption{Ground-state energy surfaces in the $\beta–\gamma$ plane for positive-parity states along the isotopic chain. White dots indicate the minimum of the energy surfaces. The energy scale varies across panels, with blue representing the minimum and red the maximum, divided into $40$ contour levels.}
    \label{fig-triaxial-pos}
\end{figure}


This evolution of the deformation can be visualized more clearly in Fig.~\ref{fig-2D-energies-pos}, which shows the axial energy curves as a function of $K^\pi$, together with the unmixed regular and intruder reference boson lines. For the lighter isotopes, the spectrum is organized into two well-defined groups: the lower-lying states correspond mainly to regular spherical configurations, while the higher states are dominated by the intruder components with a small deformation that starts developing. As neutrons are added, the two sets of configurations approach each other, interact, and eventually cross. In $^{93-99}$Nb, the ground state is spherical and the curves for the different $K^\pi$ values are almost degenerated. In the heaviest isotopes ($^{101-103}$Nb), the lowest states for each $K^\pi$ develop pronounced minima associated with intruder configurations, corresponding the lowest one to $K^\pi=5/2^+$, followed of $K^\pi=3/2^+$ and $K^\pi=1/2^+$. Note that the observed minima could be not real once the $\gamma$ degree of freedom is considered. It is also very relevant that never two minima coexist in the same energy curve.

Further insight into the shape evolution is obtained by examining the energy surfaces in the $\beta–\gamma$ plane. Fig.~\ref{fig-triaxial-pos} shows the evolution of the ground-state surface along the isotopic chain, while a more extended set of results including excited states is provided in Fig.~\ref{fig-all-pos-plane} in the Appendix \ref{sec:appendix}. The calculations reveal a smooth but clear structural evolution: the lighter isotopes ($A=93–97$) are nearly spherical, the nucleus with $A=99$ develops a modest axial deformation, and the heaviest isotopes ($A=101–103$) display pronounced triaxial minima located near $\gamma \approx 20^\circ$. 
Altogether, the results indicate that the interplay between regular and intruder configurations drives the onset of deformation but the emergence of triaxial shapes at the end of the isotopic chain is really driven by the presence of the unpaired proton, as shown in Fig.~\ref{fig-energies-betas-gammas-pos}(c) which corresponds to a calculation with $V_{mix}=0$. The analysis of the complete set of energy surfaces in Fig.~\ref{fig-all-pos-plane} in the Appendix confirms that all the cases for $A=93-97$ have essentially a spherical shape although the highest states present a more broader and flat minimum. In the case of $A=99-103$, the lowest states present a deformed shape while the highest ones rather spherical. 

\subsection{Negative parity states}
\label{sec:negative}
We now focus on the negative-parity states, which correspond to the case in which the odd proton occupies one of the $1f_{5/2}$, $2p_{3/2}$, or $2p_{1/2}$ orbits. In this case the matrix to be diagonalized has dimension $12$. 
\begin{figure}[hbt]
    \centering
    \includegraphics[width=0.5\textwidth]{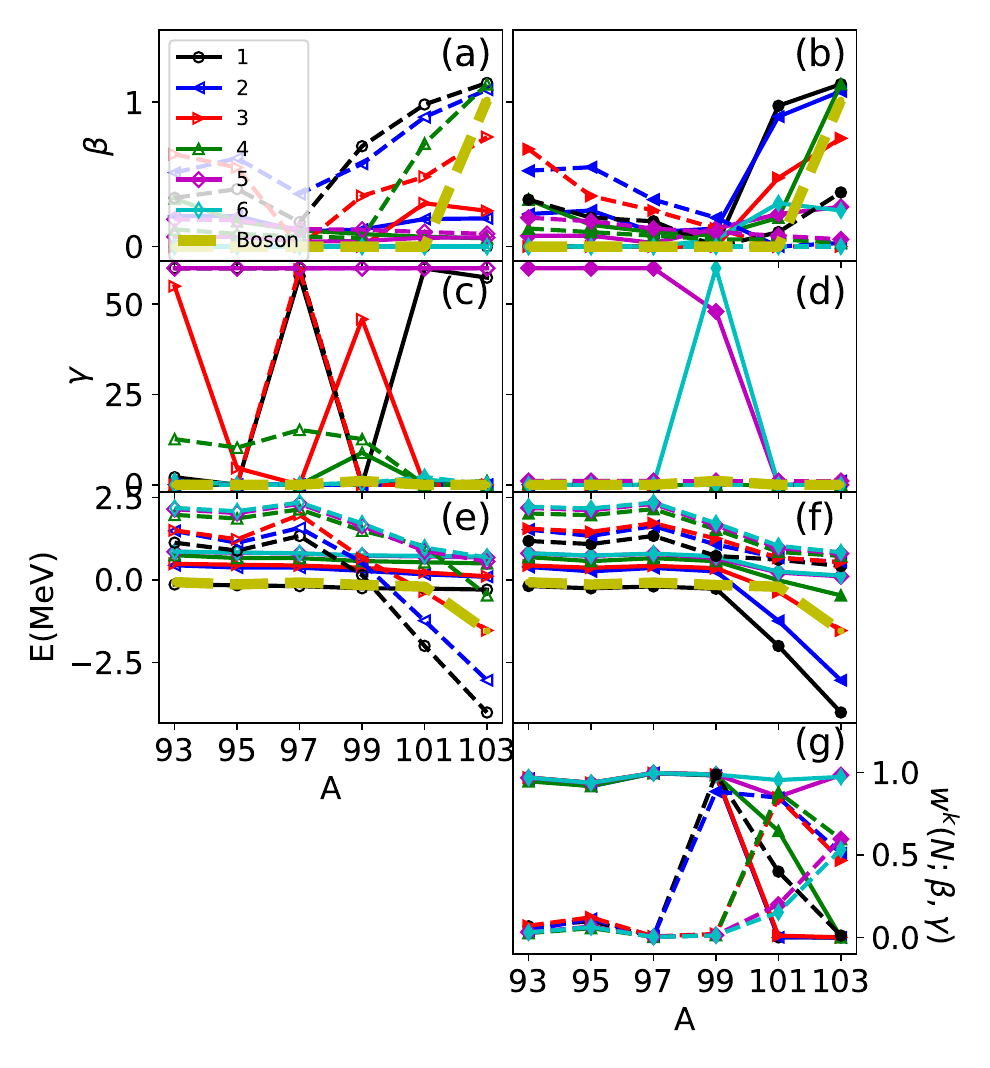}
    \caption{Same as Fig.~\ref{fig-energies-betas-gammas-pos}  but for negative-parity states.}
    \label{fig-energies-betas-gammas-neg}
\end{figure}

The first step is, once more, determining the equilibrium value of the deformation parameters and the energy for the cases without ($V_{\rm mix}=0$) and with configuration mixing included ($V_{\rm mix}\neq 0$). The main results are summarized in Fig.~\ref{fig-energies-betas-gammas-neg}. The left column corresponds to the unmixed calculation, while the right column shows the results when mixing is present. Solid lines represent regular states, whereas dashed lines denote intruder states. The thick dashed yellow lines correspond to the IBM-CM results which are provided as reference. The key point extracted from this representation is the observed behavior of the $\beta$ deformation [panel (a)] in the heavier isotopes ($^{99-103}$Nb), where different low-lying states exhibit distinct degrees of deformation, while, the lightest isotopes present a spherical shape or a small deformation in the majority of the states. This is closely related to the systematics of the excitation energies [panel (e)]: in the unmixed case, the crossing between regular and intruder configurations signals the onset of a structural change. In the case of the $\gamma$ deformation parameter, the observed oscillation is purely artificial because all situations with $\gamma\neq 0$ correspond to $\beta\approx 0$.

The case with $V_{\rm mix}\neq 0$ presents a similar behavior, but the $\beta$ deformation in the lightest isotopes is a little higher than for the case with $V_{\rm mix}=0$.  The six lowest eigenvalues are now plotted with solid lines, while the next six are shown with dashed lines. The analysis of the regular content further reveals that, for the lighter isotopes, regular and intruder states remain well separated and can be tracked individually. In contrast, for the heavier isotopes, the wave functions acquire a mixed character, indicating stronger configuration mixing. The obtained value of the regular content can be compared with the laboratory frame calculation presented in Fig.~19b of Ref.~\cite{Gavr2023}. In that figure, the ground state regular content is around $0.8$ for $^{93-95}$Nb, 1 for $^{97}$Nb, $0.9$ for $^{99}$Nb, and, finally, it changes to a value around 0 for $^{101-101}$Nb, to be compared with values around 1 for $^{93-99}$Nb and 0 for $^{101-103}$Nb in the present work. For the other states the comparison is not so easy, but it is clear that the observed transitions in the laboratory frame appears in $^{99}$Nb instead of in $^{101}$Nb.
\begin{figure}[hbt]
    \centering
    \includegraphics[width=0.5\textwidth]{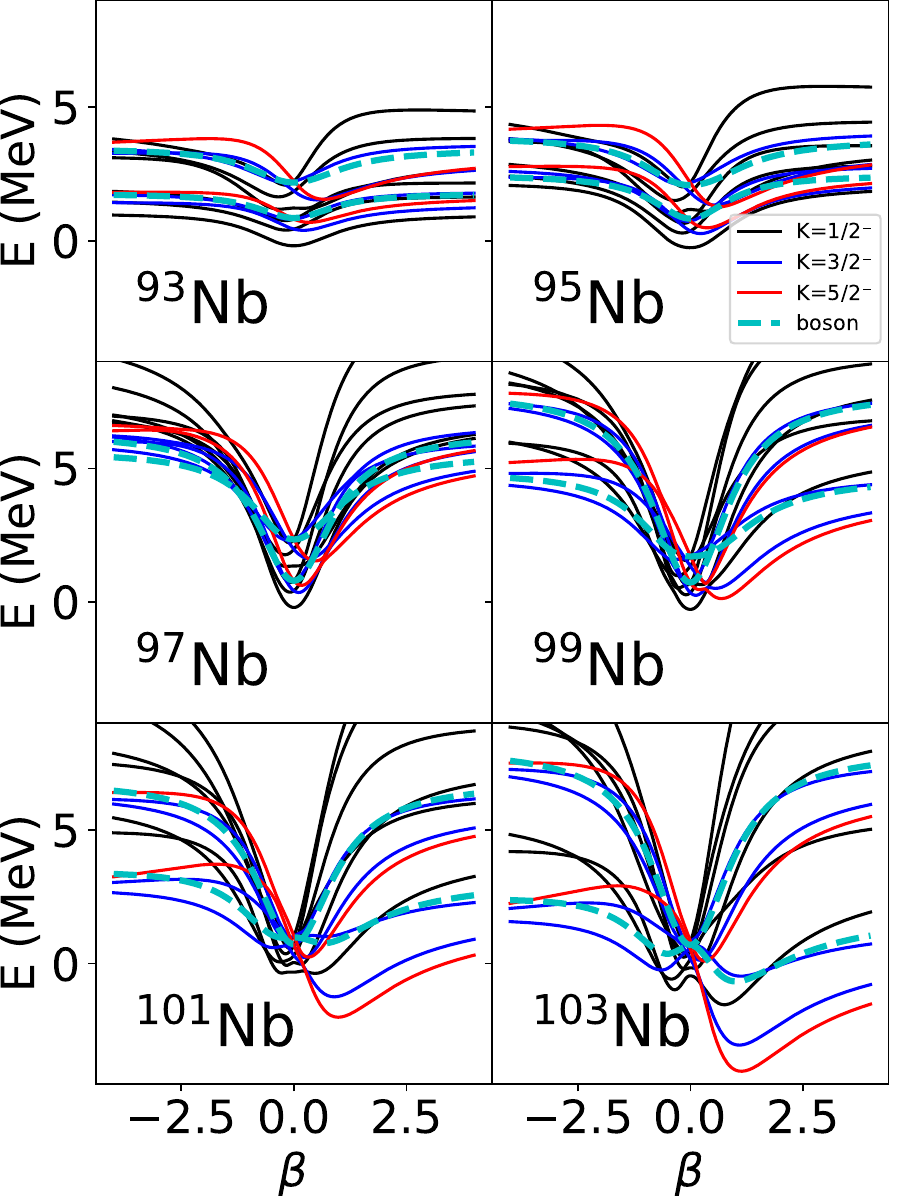}
    \caption{Same as Fig.~\ref{fig-2D-energies-pos} but for negative-parity states.}
    \label{fig-2D-energies-neg}
\end{figure}

Additional insight into the shape evolution is provided by the axial energy curves shown in Fig.~\ref{fig-2D-energies-neg}. 
For the lightest isotopes, the energy surfaces for the ground state exhibit a nearly spherical minimum for $K^\pi=\tfrac{1}{2}^-$, accompanied by another secondary minimum for $K^\pi=\tfrac{5}{2}^-$. As neutrons are added, the $K^\pi=\tfrac{5}{2}^-$ minimum increases its deformation and becomes the ground state, particularly for $^{101-103}$Nb, indicating a progressive stabilization of the prolate shape. Indeed, for these nuclei, the curves with  $K=\tfrac{3}{2}^-$ and $\tfrac{1}{2}^-$ also present well deformed minima. Note that this situation is identical for the positive-parity case. This behavior provides additional evidence for the configuration-crossing observed in the excitation spectra. However, it should be noted that these minima never coexist simultaneously in a single energy curve, instead, one minimum gradually replaces the other as the dominant configuration. 
\begin{figure}[hbt]
    \centering
    \includegraphics[width=0.5\textwidth]{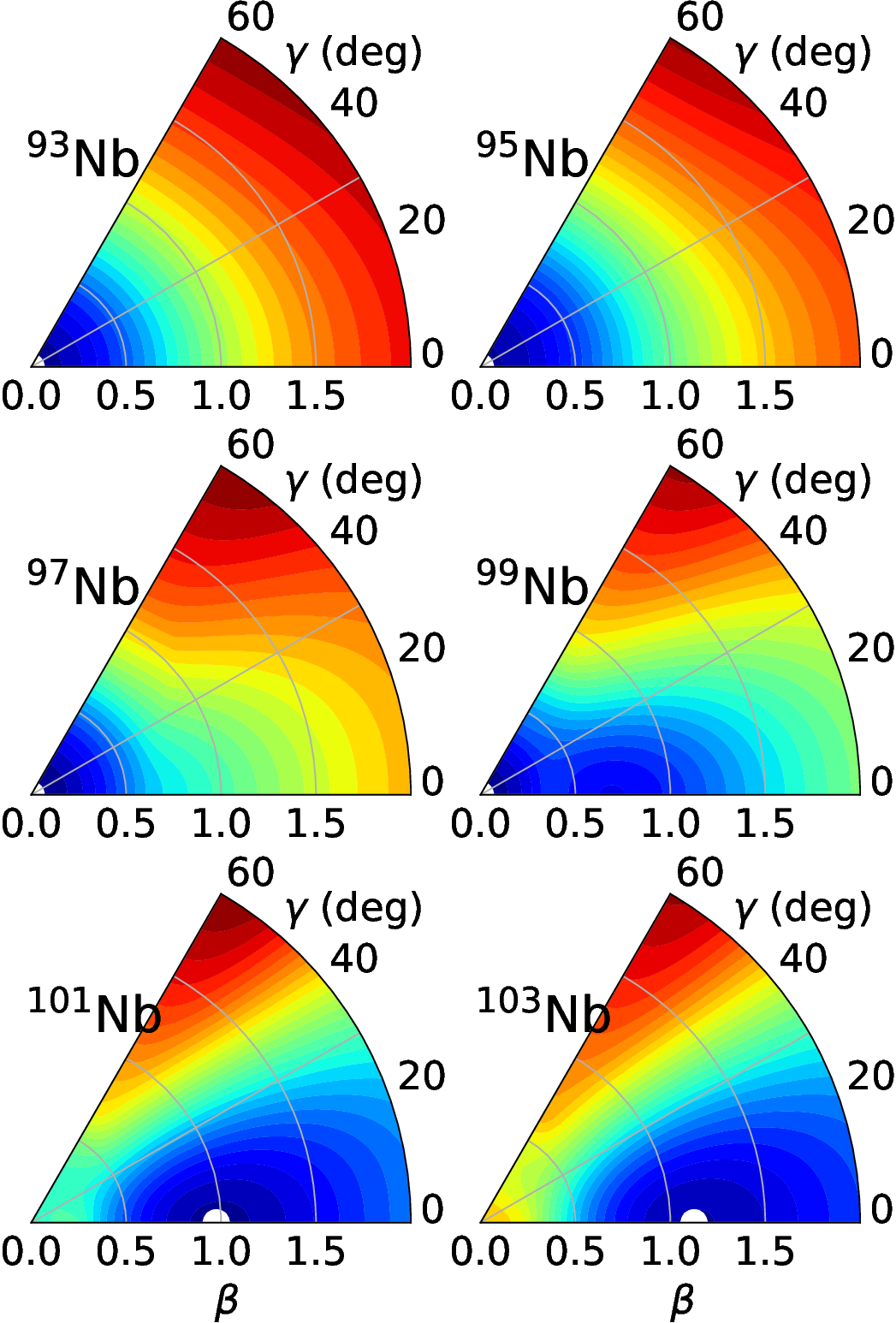}
\caption{Same as Fig.~\ref{fig-triaxial-pos} but for negative-parity states.}
    \label{fig-triaxial-neg}
\end{figure}

A more complete picture on the structural evolution can be obtained from the full energy surfaces in the $\beta-\gamma$ plane, shown in Fig.~\ref{fig-triaxial-neg}. In contrast to the positive-parity case, no triaxial minima are found for negative parity. Instead, the heavier isotopes develop well-defined prolate minima, in agreement with the axial energy curves. This shows that, although the interplay between regular and intruder configurations governs the structure in both parity sectors, the resulting shapes differ: positive-parity states evolve showing triaxiality at the end of the isotopic chain, whereas negative-parity states stabilize into axially symmetric prolate configurations. In Fig.~\ref{fig-all-neg-plane} in the Appendix, the $\beta-\gamma$ energy surfaces for all the states and isotopes are depicted. It can be observed that no triaxial minima exist, and most of the energy minima are spherical, except for the higher-energy surfaces in the lightest isotopes and the lower-energy ones in the heaviest isotopes.
\begin{figure}[hbt]
    \centering
    \includegraphics[width=0.35\textwidth]{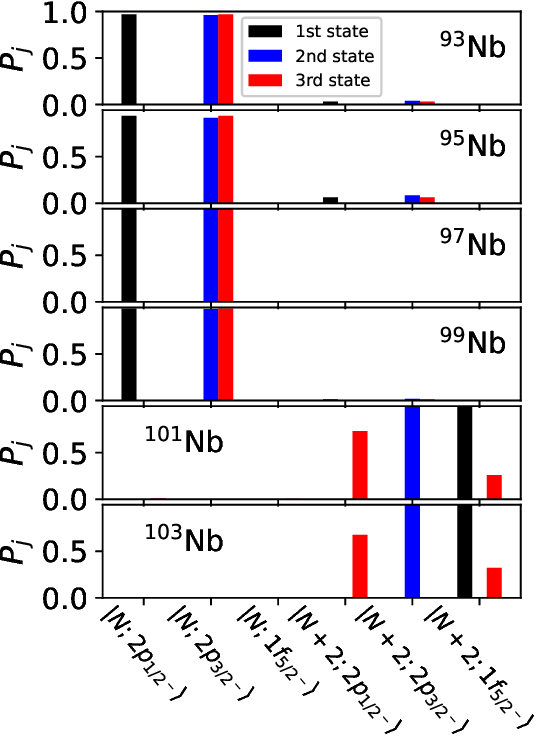}
    \caption{Single-particle occupation probability for the three first states with negative parity along the isotopic chain.}
    \label{fig-j-occupation}
\end{figure}

In this case, because the fermion can occupy different j-shells, it is possible to compute the probability occupancy. In Fig.~\ref{fig-j-occupation}, the single-particle occupation probability is depicted for the three first states. In the case of $^{93-99}$Nb, the first state presents a $1.0$ probability for the $p_{1/2^-}$ regular level, while the second and the third have almost $100\%$ probability for the $p_{3/2^-}$ regular level. In the case of $^{101-103}$Nb, the first state has a $100\%$ probability for the $f_{5/2^-}$ intruder level, while the second for the $p_{3/2^-}$ intruder one. The third state presents probabilities of around $0.6$ and $0.4$ in the levels $p_{1/2^-}$ and $f_{5/2^-}$, respectively. These results can be compared with \cite{Gavr2023}. Regarding the lowest negative parity state ($J=1/2^-$ for $^{93-99}$Nb and  $J=5/2^-$ for $^{101-103}$Nb), the agreement is good. In $^{93-97}$Nb, the second state has a clear correspondence, but not the third one, which in Ref.~\cite{Gavr2023} occupies the $f_{5/2^-}$ single-particle level, while $p_{3/2^-}$ in the present one. For $^{99}$Nb, in Ref.~\cite{Gavr2023} the second and third states present large probabilities in the intruder sector, while regular in the present work. Finally, in $^{101-103}$Nb, the agreement for the second and third states is only qualitative. 

\section{Discussion: Quantum phase transitions and excited-state energy curves}
\label{sec:qpt}
The concept of QPT has been extensively studied in the context of the IBM and its configuration mixing extension (IBM-CM) \cite{Cejn09,Hell07,Hell09}. In particular, Gavrielov \textit{et al.} recently introduced a distinction between Type I and Type II QPTs \cite{Gavr19}. In the former case, the QPT is associated with a single configuration, whereas in the latter it arises from the interplay between two different configurations. Moreover, both types of QPTs can be interconnected in certain situations, referred to as intertwined QPTs. The first documented example of this phenomenon was found in even-even Zr isotopes \cite{Gavr19}, but it has also been observed in Sr \cite{Maya2022} and Mo isotopes \cite{Maya2023}.

The situation in the case of Nb, as previously described in \cite{Gavr2023}, corresponds to an intertwined QPT. However, that study presented only the laboratory-frame perspective, i.e., no mean-field results were provided. The Nb case is very similar to that of the Zr isotopes, which constitutes its boson core. According to Figs.~\ref{fig-energies-betas-gammas-pos}(e) and \ref{fig-energies-betas-gammas-pos}(f) for positive parity [Figs.~\ref{fig-energies-betas-gammas-neg}(e) and \ref{fig-energies-betas-gammas-neg}(f) for negative parity], a clear crossing between the regular and intruder configurations occurs between $A=99$ and $A=101$. This crossing is accompanied by a sudden increase in the deformation parameter, as shown in panels (a) and (b) (and also in Fig.~\ref{fig-energies-betas-gammas-neg} for negative-parity states). It is evident that, for the ground state, the system undergoes a Type II QPT, which appears at a lower mass number $A$ than in the Zr case (yellow line). On the other hand, the intruder configuration also exhibits a pronounced evolution from an almost spherical to a deformed shape [see panel (a) of Figs.~\ref{fig-energies-betas-gammas-pos} and \ref{fig-energies-betas-gammas-neg}]. The transition is clearly more rapid for the positive-parity states than for the negative ones. Hence, one can conclude that a Type I QPT develops within the intruder configuration, and therefore Type I and Type II QPTs are intertwined in the Nb isotopes, as in the Zr case.

The obtained intrinsic-state results presented in this work for the odd-even system go somewhat beyond the conventional mean-field approach because, in fact, they provide information for the different single-particle orbits as well as for both the regular and intruder configurations. In other words, the results are not limited to the ground state of the system. For each orbit, a distinct equilibrium value of $\beta$ and $\gamma$ is obtained, which has the side effect that the resulting states lose their orthogonal character. Furthermore, the presence of two configurations gives rise to two families of orbits with markedly different deformations.  

In the case of bosonic systems, almost no attention has been paid to the highest eigenvalue obtained from the diagonalization, mainly because its physical interpretation remained unclear. In principle, one would expect a clear correspondence between the unperturbed results and those from the full diagonalization; however, this is not always the case. A detailed analysis indicates that this discrepancy arises when the unperturbed intruder configuration appears as a second local minimum of the lowest eigenenergy surface. Consequently, the absolute minimum associated with the highest eigenvalue is only weakly connected to the unperturbed intruder configuration and is instead largely influenced by the artificial minimum that emerges from the crossing of the two configurations.

In the obtained results for Nb, both families of eigenvalues appear to exhibit a clear correspondence with the unperturbed results. Consequently, the interpretation of the energy curves is relatively straightforward, including both the lowest and highest eigenvalues. For bosonic systems, this type of analysis has been carried out only for the even-even Pt isotopes in \cite{Mora08}; in general, however, only the lowest eigenvalue is usually considered \cite{Frank06,Hell07,Hell09}. A global view of all the $\beta$–$\gamma$ energy surfaces is presented in Figs.~\ref{fig-all-pos-plane} and \ref{fig-all-neg-plane} in the Appendix.

\section{Conclusions}
\label{sec:conclusions}
In this work, the odd-even isotopic chain $^{93\text{--}103}$Nb has been analyzed within the framework of the IBFM-CM intrinsic-frame formalism, using previously determined Hamiltonian parameters \cite{Gavr22b,Gavr2023}. The intrinsic-state formalism has been recently introduced in Refs.~\cite{Maya2025,Levi2025,Maya2025b}.

For positive-parity states, associated with the occupation of the $1g_{9/2}$ orbit, the results reveal a clear crossing between the regular and intruder configurations. In the absence of configuration mixing, both configurations remain well separated in energy and deformation, exhibiting a gradual crossing as the neutron number increases. When the interaction between the two configurations is taken into account, the corresponding wave functions begin to mix, reaching their maximum degree of mixing toward the heavier isotopes, although certain states still keep a pure character. This behavior leads to a structural change: the lighter nuclei are nearly spherical, whereas the heaviest isotopes develop a pronounced triaxial minimum. It should be noted that the appearance of the triaxial minimum is primarily due to the influence of the odd proton rather than to shape coexistence effects.

For the negative-parity states, corresponding to the occupation of the $1f_{5/2}$, $2p_{3/2}$, and $2p_{1/2}$ orbits, the structural evolution follows a slightly different pattern. Although a configuration crossing is also observed, the resulting mixing leads to prolate axial shapes in the heavier isotopes, without the emergence of triaxiality. Consequently, the energy minima evolve from nearly spherical configurations in the lighter nuclei to prolate shapes in the heaviest ones, namely $^{101-103}$Nb.

Finally, the results demonstrate that the interplay between the regular and intruder configurations plays a crucial role in driving the structural evolution along the $^{93-103}$Nb isotopic chain, providing a consistent explanation for the onset of deformation. In both positive- and negative-parity states, a Type II QPT is observed for the ground state, while the intruder configuration also undergoes an evolution from spherical to deformed shapes, proceeding more rapidly in positive parity and more slowly in negative parity. These findings highlight the capability of the IBFM-CM intrinsic-frame formalism to describe the shape evolution of an isotopic chain using realistic Hamiltonian parameters.

\bibliography{references-IBM-CM,references-QPT}
\newpage 
\appendix* 
\section{Complete set of $\beta-\gamma$ energy surfaces}
\label{sec:appendix}
\begin{figure*}[hbt]
    \centering
    \includegraphics[width=1.\textwidth, angle=-90 ]{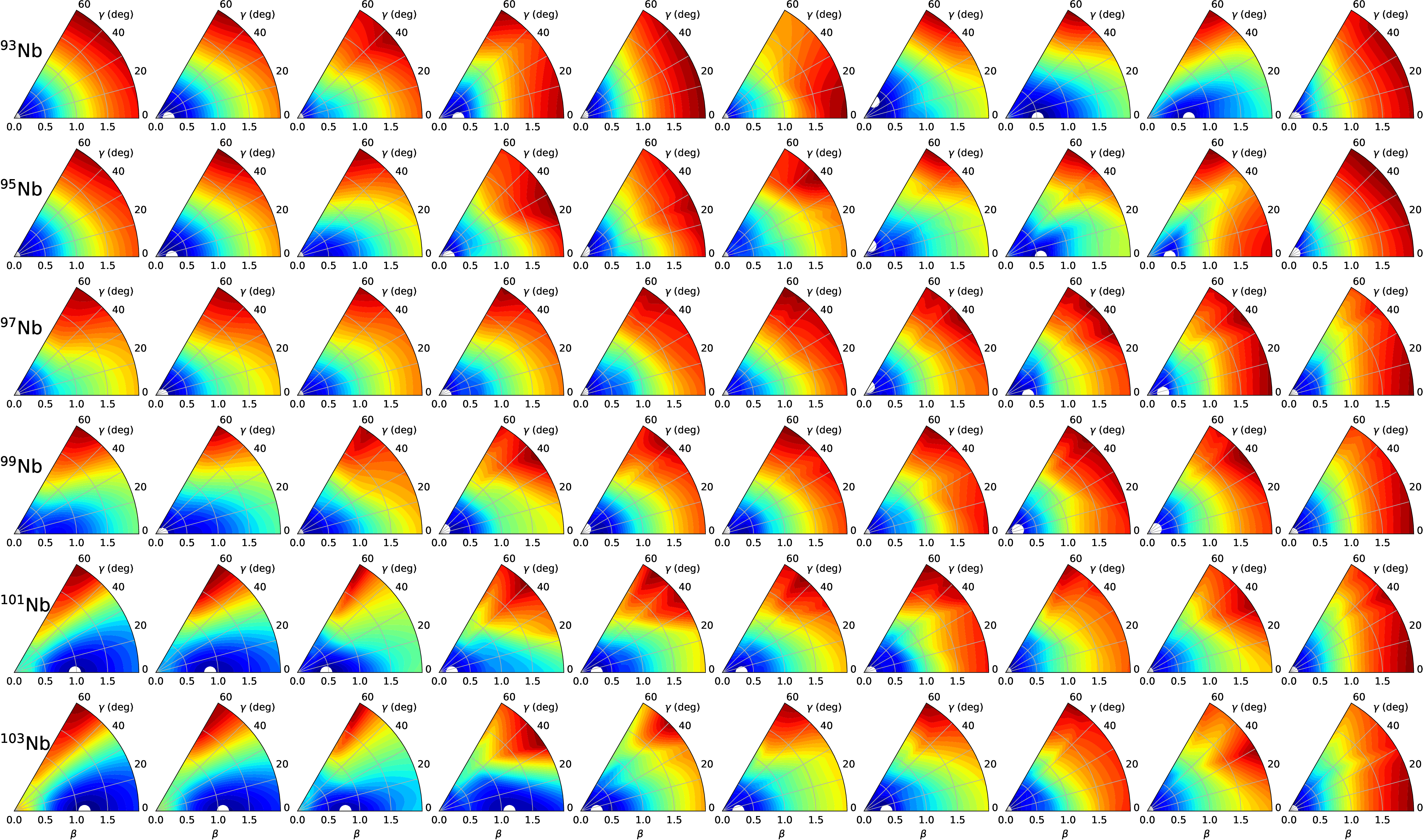}
    \caption{Energy surfaces in the $\beta–\gamma$ plane for positive-parity states across the isotopic chain. Rows correspond to the isotope while column to the ordering. White dots indicate the minimum of the energy surfaces. The energy scale varies among the panels, blue color for the minimum and red for the maximum with $40$ divisions among them.}
    \label{fig-all-pos-plane}
\end{figure*}

\begin{figure*}[hbt]
    \centering
    \includegraphics[width=1.2\textwidth, angle=-90 ]{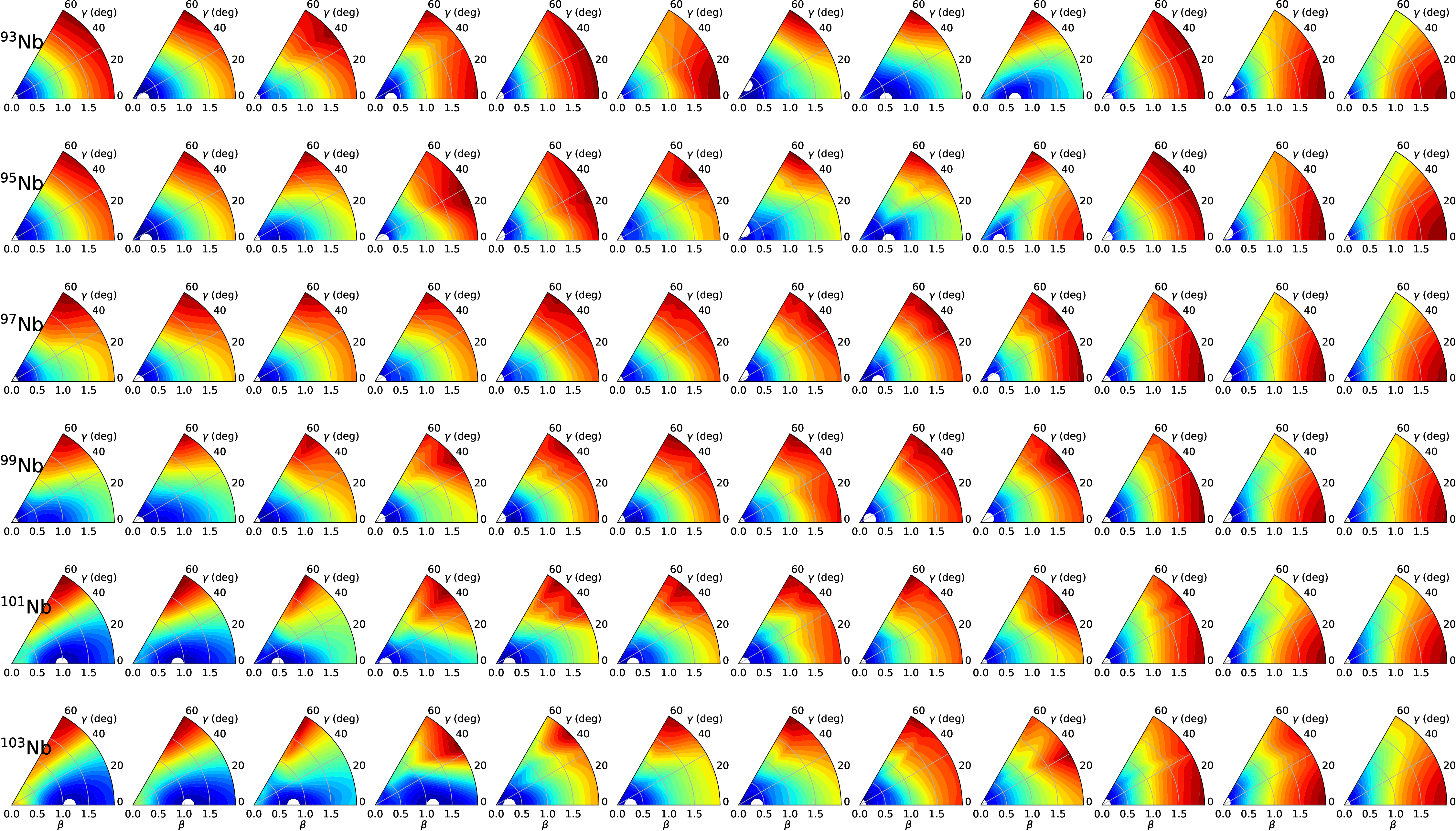}
    \caption{Same as Fig.~\ref{fig-all-pos-plane} but for negative-parity states.}
    \label{fig-all-neg-plane}
\end{figure*}

\end{document}